\documentclass[prl,twocolumn,showpacs,superscriptaddress]{revtex4}

\usepackage{amsmath,color,amssymb}
\usepackage{epsfig,graphicx,mathcomp}
\usepackage{rotating}

\usepackage{eso-pic}
\usepackage{type1cm}
\makeatletter
  \AddToShipoutPicture{%
    \setlength{\@tempdimb}{0.5\paperwidth}%
    \setlength{\@tempdimc}{0.5\paperheight}%
    \setlength{\unitlength}{1pt}%
    \put(\strip@pt\@tempdimb,\strip@pt\@tempdimc){%

    }
}

\begin{document}
\title{Modeling the Auxetic Transition for Carbon Nanotube Sheets}

\author{V. R. Coluci}\email[\footnotesize{Author to whom correspondence should be addressed. FAX:+55-19-21133364. Electronic address: }]{vitor@ceset.unicamp.br}
\affiliation{Center for High Education on Technology, 
             University of Campinas - UNICAMP 13484-332, Limeira,
             SP, Brazil}

\affiliation{Applied Physics Department, Institute of Physics P.O.Box 6165,
             University of Campinas - UNICAMP 13083-970, Campinas,
             SP, Brazil}

\author{L. J. Hall}
\affiliation{Alan G. MacDiarmid NanoTech Institute, University of Texas at Dallas, Richardson, 
TX 75083-0688, USA}

\author{M. E. Kozlov}
\affiliation{Alan G. MacDiarmid NanoTech Institute, University of Texas at Dallas, Richardson, 
TX 75083-0688, USA}

\author{M. Zhang}
\affiliation{Alan G. MacDiarmid NanoTech Institute, University of Texas at Dallas, Richardson, 
TX 75083-0688, USA}

\author{S. O. Dantas}
\affiliation{Departamento de F\'{\i}sica, Universidade Federal de Juiz de Fora, 36036-330 Juiz de Fora MG, Brazil}

\author{D. S. Galv\~ao}
\affiliation{Applied Physics Department, Institute of Physics P.O.Box 6165,
             University of Campinas - UNICAMP 13083-970, Campinas,
             SP, Brazil}

\author{R. H. Baughman}
\affiliation{Alan G. MacDiarmid NanoTech Institute, University of Texas at Dallas, Richardson, 
TX 75083-0688, USA}

\date{\today}

\begin{abstract}
A simple model is developed to predict the complex mechanical properties of carbon nanotube sheets (buckypaper) [Hall \textit{et al.}, \textit{Science} \textbf{320} 504 (2008)]. Fabricated using a similar method to that deployed for making writing paper, these buckypapers can have in-plane Poisson's ratios changed from positive to negative, becoming auxetic, as multiwalled carbon nanotubes are increasingly mixed with single-walled carbon nanotubes. Essential structural features of the buckypapers are incorporated into the model: isotropic in-plane mechanical properties, nanotubes preferentially oriented in the sheet plane, and freedom to undergo stress-induced elongation by both angle and length changes. The expressions derived for the Poisson's ratios enabled quantitative prediction of both observed properties and remarkable new properties obtainable by structural modification.
\end{abstract}

\maketitle

\section{I. Introduction}
The ratio of percent lateral contraction to percent applied tensile elongation is the Poisson's ratio. If the lateral dimension expands during stretching, the Poisson's ratio is negative and the material is called auxetic \cite{lakes}. Recent interest in this counter-intuitive behavior originated from pioneering discoveries that partially collapsed foams and honeycombs \cite{lakes,gibson}, fibrillar polymers \cite{evans}, and polymer composites \cite{milton} can be auxetic. This unusual property of auxetic materials results in various useful effects, such as increased indentation resistance and increased shear stiffness \cite{2}. Possible or deployed applications of auxetic materials are, for example, anti-ballistic vests, air filters, strain sensors, molecular-scale amplifiers, vascular implants, gaskets, sound absorbers, artificial muscles, and wrestling mats \cite{lakes,2,baughman}. Due to their unusual and interesting properties, auxetic materials have been the subject of intense experimental and theoretical research \cite{2,3,doug}.

Recently, we showed that the Poisson's ratio of carbon nanotube sheets (buckypaper) can change from positive to negative as multiwalled carbon nanotubes (MWNTs) are increasingly mixed with single-walled carbon nanotubes (SWNTs) \cite{hall}. While the in-plane Poisson's ratio for SWNT nanotube sheets was positive (about 0.06) and slightly changes until MWNT content reached 73 weight percent (wt. \%), further addition of MWNTs decreased Poisson's ratio to $-$0.20 (Fig. 1). On the other hand, large positive Poisson's ratios were observed for the thickness direction: 0.33$\pm$0.14 and 0.75$\pm$0.30 for SWNT and MWNT sheets, respectively. A non-linear dependence of Young's modulus, strength, and toughness on MWNT content was also observed, though electronic conductivity and density depended approximately linearly on MWNT content. A model incorporating an idealization of the complex structure of the SWNT/MWNT buckypapers and the main deformation mechanisms was proposed in order to understand the behavior of the in-plane and thickness-direction Poisson's ratios within these nanotube sheets \cite{hall}.  
We herein provide a full account of calculation methods and results, which were previously briefly outlined \cite{hall}.

The Poisson's ratios of individual SWNTs and MWNTs have been theoretically obtained using analytical \cite{popov,shen,mintmire} and atomistic models including empirical potentials \cite{elastic,lu}, tight-binding-based approaches \cite{hernandez}, and \textit{ab initio} methods \cite{portal}. All of these investigations predict positive values for the Poisson's ratios. Calculations using density functional theory \cite{portal} provide values between 0.12 and 0.19, comparable to the Poisson's ratio for the basal plane in graphite (0.16) \cite{vgraf1,vgraf2}. However, when individual nanotubes are assembled together in sheets containing fiber networks, either negative or positive Poisson's ratios can arise \cite{hall}. Analyzing carbon nanotube sheets, Berhan \textit{et al.} \cite{berhan} showed using Euler beam-network simulations that increasing the number of interfiber connections can lead to improvements in carbon nanotube sheet stiffness.

Fiber networks have been subject of extensive investigation over the last six decades (e.g. \cite{cox,warren,narter,astrom,wu,peter}). Using effective-medium theory, Cox's pioneer work on fibers \cite{cox} predicts the effective moduli of two-dimensional fibers considering only the fiber axial deformation. Analyzing a solid mat of fibers, Cox predicted the possibility of a negative Poisson's ratio in plane of the paper existing concurrently with a high positive value through the thickness of the paper.  However, the Poisson's ratio measured in-plane for ordinary fiber networks, like writing paper, are rather large and positive. Extensions of the Cox's model have been developed where fiber bending, elongation, and contraction, as well as Poissonian distribution of fiber segment lengths, are taken into account \cite{warren,narter,astrom,wu}. Producing carbon nanotube networks from dispersed aqueous nanotube suspensions \cite{hall}, we were able to provide an experimental realization of Cox's prediction that a negative Poisson's ratio can exist for paper-like fiber mats. The present goal is to describe a simple model that provides a realistic, though simplified, description of the structural nature of buckypaper, and use this model to explain why the observed Poisson's ratio of multiwalled and single walled carbon nanotube buckpaper differ in sign.

This paper is divided into the following: Section II briefly describes the experimental details of the buckypaper fabrication and the methods used to determine the Poisson's ratios. The model and the derivation of the Poisson's ratios are presented in Sec. III. In Sec. IV, we present the results obtained from the model and  discuss the main results in Sec. V. Conclusions and final remarks of this work are summarized in Sec. VI.

\begin{center}
\begin{figure}[ht]
\includegraphics[angle=0,scale=0.3]{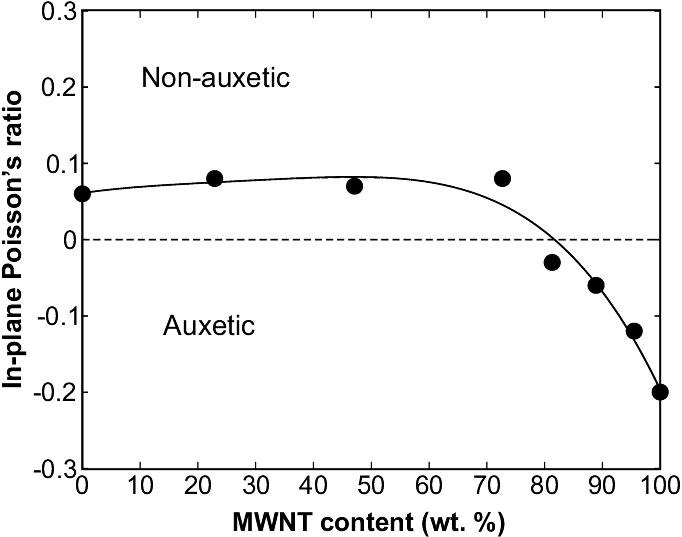}
\caption{Measured in-plane Poisson's ratio vs. MWNT content in SWNT/MWNT sheets. The continuous line is only a guide to the eye.}
\end{figure}
\end{center}

\vspace{-2cm}
\section{II. Experimental Details}
The utilized MWNTs ($\sim$12 nm in diameter, $\sim$200 $\mu$m long, and containing about 9 walls) were harvested from nanotube forests prepared by chemical vapor deposition (CVD) of acetylene gas \cite{zhang}. The SWNTs ($\sim$1.0 nm diameter and below micron long) \cite{yudasaka,bachilo} were commercially produced by Carbon Nanotechnologies, Inc. using the HiPco synthesis method by high pressure CVD of carbon monoxide \cite{nikolaev}. The MWNTs had below 2 wt. \% catalyst while the unpurified HiPco nanotubes have high wt. \% catalyst, though low volume \% catalyst. Catalyst concentration is apparently unimportant for the used as-synthesized HiPco SWNTs, since buckypaper sheets made from commercially obtained as-synthesized and acid-reflux purified SWNTs (Buckypearls) differed in in-plane Poisson's ratio by only 0.034.
The nanotube buckypaper was fabricated by vacuum filtration of an ultrasonically dispersed aqueous nanotube suspension containing Triton X-100 surfactant, washed with successively water and methanol, vacuum drying (85$^\circ$C for two days), and then peeling the nanotube sheet from the filter \cite{rinzler}. Nanotube sheet thickness was held approximately constant (50 $\mu$m). Scanning electron microscopy indicated that SWNTs and MWNTs are intimately commingled in sheets comprising both nanotube types. The average angle between the nanotube length direction and the sheet plane (41.7$^\circ$ for MWNT sheets and 45.0$^\circ$ for SWNT sheets) was determined by diffraction for incident x-rays in an in-plane sheet direction, using the dependence of diffraction intensities on azimuthal angle \cite{ran}. Fig. 2 shows scanning electron microscopy images of buckypaper surfaces for sheets produced with different MWNT wt. \% contents. 

Reported in-plane mechanical properties measurements are for 2 mm $\times$ 12 mm carbon nanotube strips during deformation at 0.10\% strain/minute. Poisson's ratio measurements utilized nanotube sheets coated with trace TiO$_2$ particles for marking position. Digital images were captured during constant rate tensile deformation, and interpreted using image correlation software (Vic-2D Correlated Solutions, Inc., West Columbia SC, USA) to obtain changes in the separations between thousands of TiO$_2$ particles as a function of tensile stress, corresponding sheet strains in stretch and lateral directions, and the Poisson's ratio. The thickness-direction Poisson's ratio was obtained from scanning electron micrographs showing sheet thickness versus applied in-plane tensile strain.

\begin{center}
\begin{figure}
\includegraphics[angle=0,scale=0.3]{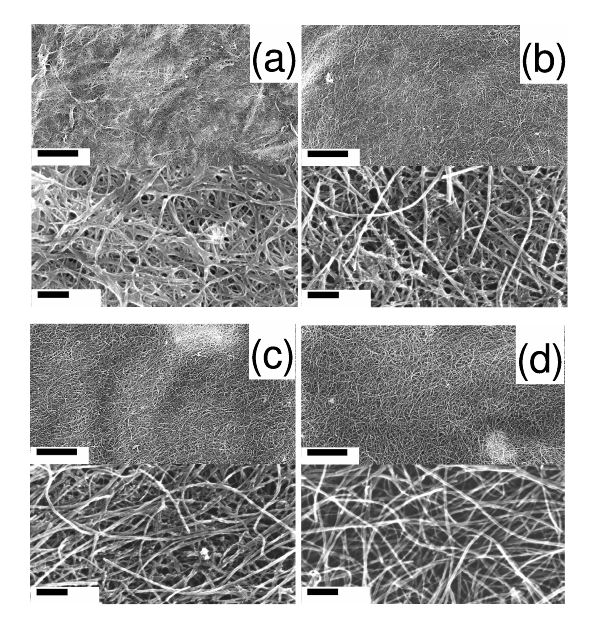}
\caption{Scanning electronic microscopy images of the surfaces of buckypaper containing (a) 0 wt. \%, (b) 47.1 wt. \%, (c) 72.7 wt. \%, and (d) 100 wt. \% MWNT content. Different magnifications are shown in top and bottom parts of each image. The scale bars for the upper and lower images in (a)-(d) correspond to 2$\mu$m and 200 nm, respectively.}
\end{figure}
\end{center}

\vspace{-1cm}
\section{III. Development of the model}
\subsection{Structural model}
As we can see from Fig. 2, the structures of the carbon nanotube sheets are very complex,  nanotubes and nanotube bundles having various diameters meander in three dimensions, like cooked spaghetti. Therefore, the challenge is to formulate a model for these intractably complex nanotube sheets that is sufficiently simple that it can be tested and used for prediction of future results. A first approach to achieve this is to simplify the complex morphology of the meandering nanotubes in the thickness direction to a zigzag set of struts as illustrated in Fig. 3 (a). The bends at the zigs and zags enable tractable representation of the observed deviation of nanotubes from perfect in-plane alignment and the geometrical effect of nanotube straightening on in-plane and thickness direction Poisson's ratios. The bend force constant at the zigs and zags correspond to the effective force constant for elongating a meandering nanotube in a network of interacting neighboring nanotubes. Coupling between intersecting nanotubes is at junctions, where the zags from one layer of zigzag chains are coupled to the zigs for the next layer of chains.

\begin{center}
\begin{figure}[b]
\includegraphics[angle=0,scale=0.3]{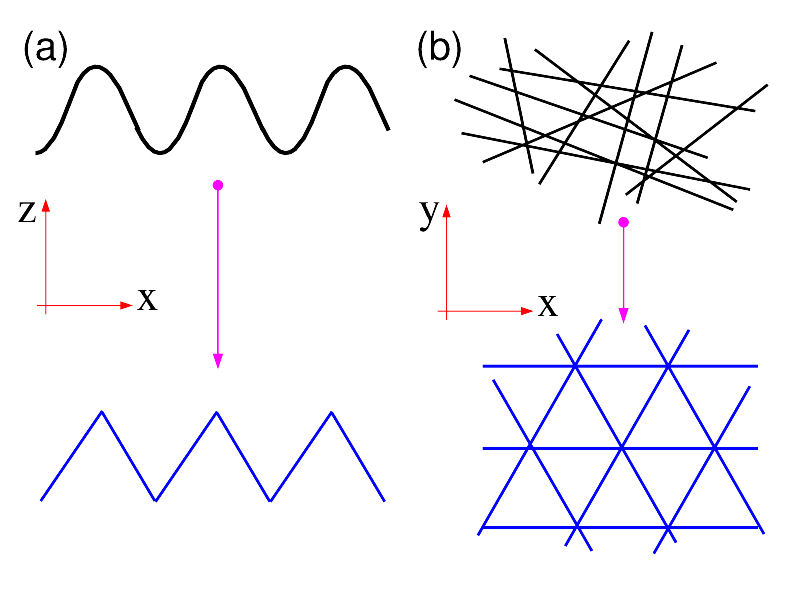}
\caption{(Color online) Schematic representation of the approximation used to derive a simple model for the nanotube sheets. (a) A SWNT bundle or a MWNT  that is deformed along the in-plane and thickness directions (top) is represented by a set of struts forming a zigzag arrangement (bottom). (b) The complex nanotube sheet morphology (top) is represented by a simplified model of an ordered structure (bottom).}
\end{figure}
\end{center}

The second simplification is to describe the disordered observed structure by an mechanically equivalent structure comprising layers that are periodic within the sheet plane as an ordered structure (Fig. 3 (b)). A simple example of an ordered structure that could be used for representing the nanotube sheets is similar to an egg-rack. This structure is composed of oppositely facing `four-legged claws' arranged on a square grid. Fig. 4 depicts a schematic view of a ``egg-rack''-type structure. Grima \textit{et al.} \cite{egg} demonstrated that when this structure is loaded in tension, the connectivity of the claws forces them to open in all directions, hence producing a negative Poisson's ratio in the plane of the structure and a positive Poisson's ratio in the thickness direction. These two features of such type of structure are observed in the real MWNT sheets. The meandering of fibers can be represented in this structural model as indicated in Fig. 4 by the magenta and red struts. Moreover, this representation also permits the fibers to cross each other, like they do in the real structure (Fig. 2).

\begin{center}
\begin{figure}
\includegraphics[angle=0,scale=0.3]{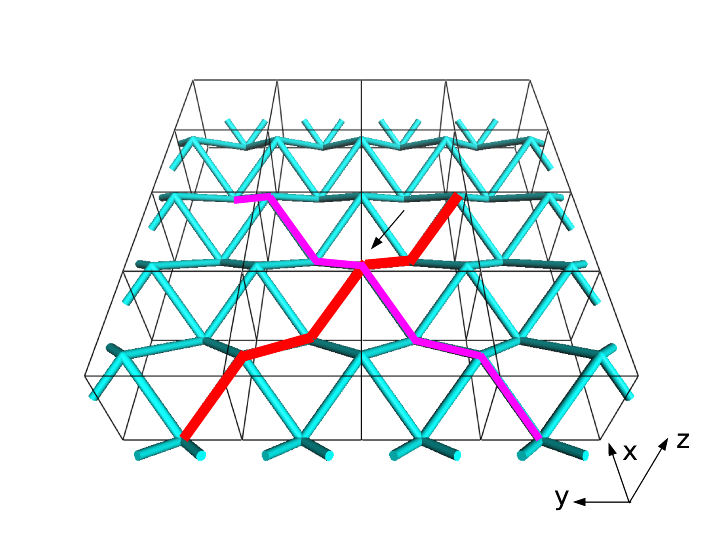}
\caption{(Color online) Schematic view of an ``egg rack''-type structure representing the real complex structure of the nanotube sheets. The arrow indicates the position where two model fibers would be in contact. Due to the connectivity of the structure, tensile loading in the in-plane direction ($xy$ plane) produces a negative Poisson's ratio in the plane of the structure and a positive Poisson's ratio in the thickness direction ($xz$ and $yz$ planes) if the only deformation made is angle bending.}
\end{figure}
\end{center}

\vspace{-1.0cm}
While the egg-rack model incorporates aspects observed in real nanotube sheets, such as preferential nanotube orientation in the sheet plane (but with positive and negative deviations from in-plane orientation) and quite different Poisson's ratios for sheet plane and sheet thickness directions, it lacks the needed isotropy for in-plane mechanical properties. Anisotropic models have been previously deployed to represent sheets that have isotropic in-plane properties \cite{delince}. Uncertainty in the sign of Poisson's ratio resulted from the need to average in-plane properties to obtain predicted properties for sheets having isotropic in-plane mechanical properties. Thus, in order to avoid the averaging process, a structural model showing in-plane isotropy is desired. 

\begin{center}
\begin{figure}
\includegraphics[angle=0,scale=0.3]{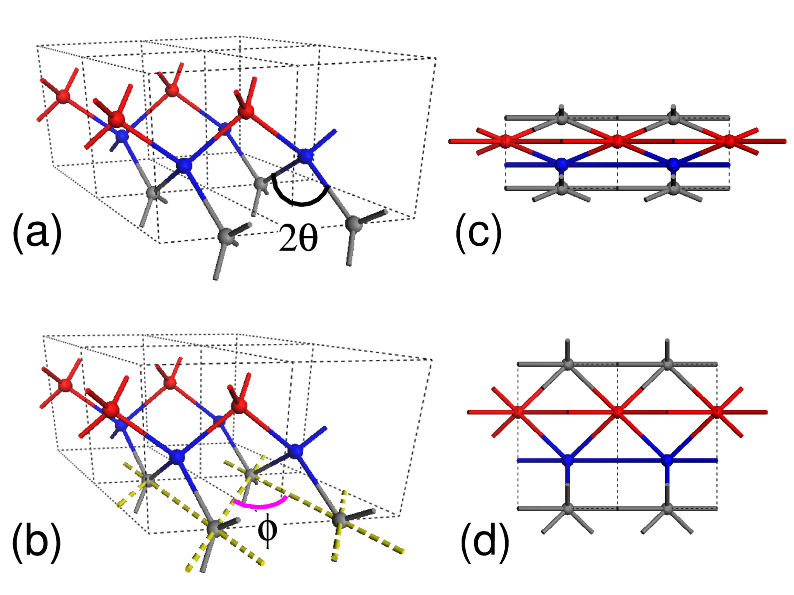}
\caption{(Color online) (a)-(b) Three-dimensional structure used to represent nanotube sheets. Each strut represents a fiber and each ball a fiber junction. In (b) the $\phi$ angle is represented through virtual struts (yellow dashed lines). (c)-(d)  Lateral views of the structure with different inclination angles ($\gamma$) of the struts with respect to the sheet plane: (c) 20$^\circ$ and (d) 40$^\circ$.}
\end{figure}
\end{center}

\begin{center}
\begin{figure}
\includegraphics[angle=0,scale=0.3]{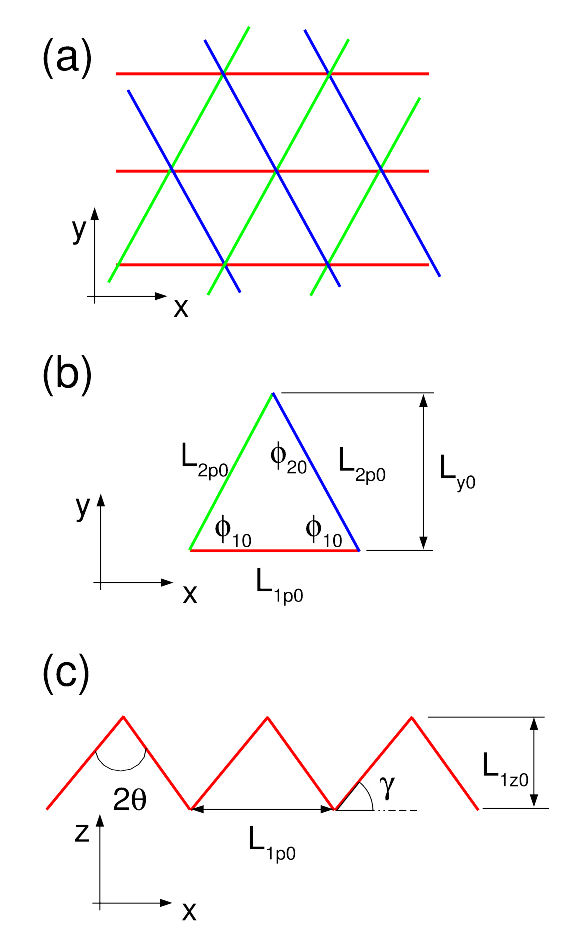}
\caption{(Color online) (a)-(b) Labeled schematic illustration of the hexagonal model structure viewed perpendicular to the sheet plane, where the zigzag nanofibers in the same layer share the same color. (c) Labeled lateral view of a zigzag nanofiber in this structural model.}
\end{figure}
\end{center}

\vspace{-1.0cm}
The simplest model that provides all these key features of the nanotube sheets has the hexagonal space group P6$_2$22 and inter-nanotube non-covalent junctions located at (0.5, 0, 0) and equivalent locations in the unit cell. In this model, the nanotubes (and nanotube bundles) are represented by zigzag chains parallel to the sheet plane (with angle between the struts and the basal plane of $\pm \gamma$ and an inter-strut angle of 2$\theta$ = $\pi - 2\gamma$). Zigzag chains in one nanotube sheet layer connect non-covalently with those in the next layers at the extremes of the zigs and zags, where torsion about the contact enables change in the intersection angle $\phi$ between nanotubes. A three-dimensional view of the structure is presented in Fig. 5. The same mechanical properties result for the closely related structure shown in Fig. 6, in which each successive layer of zigzag chains are equally likely to be added in either of two possible directions. Before deformation by application of tensile stress along the $x$-direction, struts 1 have length $L_{10}$ and length projected onto the sheet plane ($xy$ plane) of $L_{1p0}$, and struts 2 have the same length ($L_{20}$) and projected length ($L_{2p0}$) as for struts 1 (Fig. 6). For the special case of $\gamma = 0$, $L_{1p0} = L_{10}$. Thus the structural model depicted in Fig. 6 is the approximation used here to represent the complex structure of nanotube sheets.

The last step on the model definition is to decide which nanotube deformation mechanisms in the real nanotube sheets will be taken into account within the model. Due to the complex morphology of the nanotube sheets, the inclusion of all possible deformation types would not be possible for a predictive model. Therefore, we incorporated only the deformation modes we considered to be most important for describing the mechanical properties of the nanotube sheets. Thus, the following elastic deformations were considered: nanotube (or nanotube bundle) axial stretching, represented by the force constant $k_s$; nanotube bending due to changes in the $\theta$ angle, represented by the bending force constant $k_\theta$; and due to changes in the torsional angle $\phi$ between coupled intersecting nanotubes, represented by the torsional force constant $k_t$. These force constants are effective values, arising in the complex real structure from the energy needed to straighten meandering nanotubes and change the angle between intersecting nanotubes. 

Before we proceed with the determination of the expressions for the Poisson's ratios of the hexagonal model, it is useful to derive the effective force constant $k_{sb}$ of a single zigzag nanofiber when both strut stretch and $\theta$ angle deformation are included. The zigzag chain is depicted in Fig. 7. The dimensions of the unit cell are given by $L_x = 2L\sin\theta$ and $L_z = L\cos\theta$, where $L$ is the length of the strut and $2\theta$ is the inter-strut bond angle. The zigzag chain force constant is related to the applied force $F$ through $k_{sb}= 2F/\Delta L_x$.

\begin{center}
\begin{figure}[h]
\includegraphics[angle=0,scale=0.3]{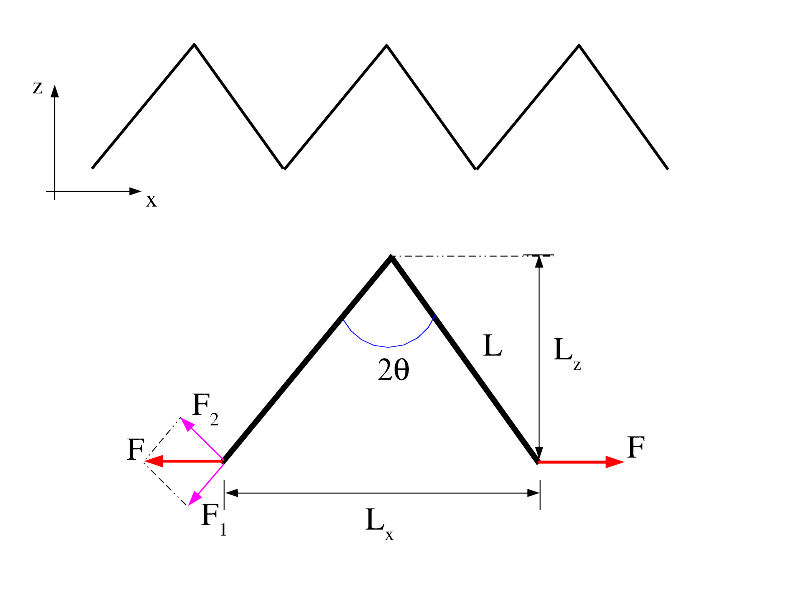}
\caption{(Color online) Zigzag chain (top) and its corresponding unit cell (bottom).}
\end{figure}
\end{center}

\vspace{-1.0cm}
The force $F$ can be written in terms of the parallel $F_1$ ($= F \sin\theta$) and perpendicular $F_2$ ($= F\cos\theta $) components with respect to the strut. The forces $F_1$ and $F_2$ will then produce the corresponding displacements $\delta_1=F_1/k_s$ and $\delta_2=F_2/k_\theta$. Thus, the variation $\Delta L_x$ can be obtained using the components of $\delta_1$ and $\delta_2$, leading to $\Delta L_x= 2F(\sin^2\theta/k_s+\cos^2\theta/k_\theta)$.  Consequently, $k_{sb}$ will be given by:
\begin{equation}
\displaystyle k_{sb}= \frac{k_s k_\theta}{k_\theta \sin^2\theta+k_s \cos^2\theta}.
\end{equation}

We can see that when $k_\theta/k_s\rightarrow 0$, $k_{sb} \rightarrow k_\theta/\cos^2\theta$ and when $k_\theta/k_s\rightarrow \infty$, $k_{sb} \rightarrow k_s/\sin^2\theta$.

\subsection{Derivation of the Poisson's ratios}
When an in-plane tensile stretch is applied along the $x$-direction, causing a small tensile strain $\epsilon = \Delta L_{1p}/L_{1p0}$ ($\ll 1$), strut lengths and angles change to $L_{1p} = L_{1p0} + \Delta L_{1p}$, $L_{2p} = L_{2p0} + \Delta L_{2p}$,  $\phi_1  = \phi_{10} + \Delta \phi_1$, $\phi_2  = \phi_{20} + \Delta \phi_2$ ($\phi_{10} = \phi_{20} = \phi_{0}=$ 60$^\circ$), and likewise for $\theta$. The total energy per strut ($E$) needed for producing a given small tensile strain in terms of angle bend and strut length changes is given by: 

\begin{eqnarray}
\displaystyle E&=&(k_{sb}/6)[(\Delta L_{1p})^2+2(\Delta L_{2p})^2]\\ \nonumber
&+&(k_{t}/6)[2( L_{1p0} \Delta\phi_{1})^2+(L_{2p0}\Delta\phi_{2})^2].
\end{eqnarray}

Minimization of $E$ provides all changes in lengths and angles for a specified small tensile strain in the in-plane direction. From these changes, the in-plane ($\nu_1$) and sheet thickness direction ($\nu_3$) Poisson's ratios can be obtained. The energy must be minimized subject to the constraint that all layers have the same tensile-direction and width-direction strains:  (i) $\phi_2+2\phi_1=\pi$ and (ii) $L_{1p}=2L_{2p}\sin(\phi_2/2)$. Due to constraint (ii)  $\Delta L_{1p}$ is written as 

\begin{equation}
\displaystyle 
\Delta L_{1p}=L_{0}\cos((\phi_0+\Delta\phi_2)/2)\Delta\phi_2 + 2 \sin(\phi_0/2)\Delta L_{2p}. 
\label{eq_deltaL}
\end{equation}
where we used $L_{1p0}=L_{2p0}=L_0$. In the elastic regime $\Delta \phi_2$ is small, therefore, $\cos((\phi_0+\Delta\phi_2)/2)\simeq\sqrt{3}/2-\Delta\phi_2/4$, thus (\ref{eq_deltaL}) can be written as 

\begin{equation}
\displaystyle \epsilon  =\left(\frac{\sqrt{3}}{2}-\frac{\Delta\phi_2}{4}\right)\Delta\phi_2 +\delta, 
\end{equation}
where  $\delta\equiv \Delta L_{2p}/L_{0}$. Since we are comparing terms up to the first order in $\delta$ and $\epsilon$ in the above equation, it is reasonable to assume that $(\Delta\phi_2)^2$ is very small when compared to $\epsilon$ and $\delta$. Therefore, the constraints (i) and (ii) are expressed as: (a) $\Delta\phi_2+2\Delta\phi_1=0$ and (b) $\epsilon =\sqrt{3} \Delta\phi_2/2 +\delta$, with $\epsilon$,$\delta\ll 1$. Using (b) the energy $E$ becomes
\begin{equation}
\displaystyle E=(k_{sb}/6)[(\epsilon L_0)^2+2(\delta L_0)^2]+(3/2)(k_{t}/6)(L_0\Delta\phi_{2})^2.
\end{equation}

Defining $f\equiv 6E/(k_t L_0^2)$ and $R\equiv k_{sb}/k_t$ the function to minimize can be expressed as 
\begin{equation}
\displaystyle \displaystyle f(\delta,\Delta\phi_{2}) = R(\epsilon^2+2\delta ^2)+(3/2)\Delta\phi_{2}^2,
\end{equation}
subject now to the following constraint:
\begin{equation}
\displaystyle g(\delta,\Delta\phi_{2}) =\epsilon -\frac{\sqrt{3}}{2} \Delta\phi_2 -\delta.
\end{equation}

The solutions can be obtained by using the Lagrange's multipliers method, solving the system:

\begin{equation}
\displaystyle g(\delta,\Delta\phi_{2})=0,\;\;\; \frac{\partial \tilde{f}(\delta,\Delta\phi_{2})}{\partial \delta}=0,\;\;\;\frac{\partial \tilde{f}(\delta,\Delta\phi_{2})}{\partial \Delta\phi_{2}}=0,
\end{equation}
where $\tilde{f}(\delta,\Delta\phi_{2})\equiv f(\delta,\Delta\phi_{2})-\lambda g(\delta,\Delta\phi_{2})$ and $\lambda$ is the Lagrange's multiplier. The solution for this system is 

\begin{equation}
\displaystyle \Delta \phi_2 =\frac{2R}{\sqrt{3}(1+R)}\epsilon
\end{equation}
and
\begin{equation}
\displaystyle  \delta =\frac{\epsilon}{1+R}.
\end{equation}

The energy at the minimum will then assume the value
\begin{equation}
\displaystyle E_{min} =\frac{R(R+3)}{6(1+R)} \;k_t L_0^2\epsilon^2.
\label{eq-energy}
\end{equation}

The in-plane Poisson's ratio $\nu_{1}$ is given by
\begin{equation}
\displaystyle \nu_{1}=-\left(\frac{\Delta  L^*/L^*}{\Delta L_{1p}/L_0} \right),
\end{equation}
where $L^*=L_{2p}\cos(\phi_2/2)$. Neglecting second order terms we can write  

\begin{equation}
\displaystyle \Delta L^*\simeq -\frac{L_0}{2}\;\sin(\phi_0/2) \Delta\phi_2+\cos(\phi_0/2)\;\Delta  L_{2p},
\end{equation}
or
\begin{equation}
\displaystyle \Delta L^* = \frac{L_0}{2}(\sqrt{3}\; \delta-\Delta\phi_2/2).
\end{equation}

Consequently:
\begin{equation}
\displaystyle \nu_{1}=\frac{R-3}{3(1+R)}.
\end{equation}

Following the definitions of Fig. 7, the Poisson's ratio for a single zigzag chain is given by $\nu_{zig}=-(\Delta L_z/L_z)/(\Delta L_x/L_x)$. The term $\Delta L_z$ is calculated using $\delta_1$ and $\delta_2$, leading to $\Delta L_z= F\sin\theta\cos\theta(1/k_s-1/k_\theta)$. Therefore,
\begin{equation}
\nu_{zig}=\frac{\tan^2\theta(k_s/k_\theta -1)}{\tan^2\theta + k_s/k_\theta}.
\end{equation}

The thickness-direction Poisson's ratio $\nu_{3}$ is determined through a three-layer average, i.e.,
\begin{equation}
\displaystyle \nu_{3}=(1/3)\left(-\frac{\Delta L_z^{(1)}/L_z^{(1)}}{\Delta L_{1p}/L_0}-2\frac{\Delta L_z^{(2)}/L_z^{(2)}}{\Delta L_{1p}/L_0}\right),
\end{equation}
where $\Delta L_z^{(i)}/L_z^{(i)}$ refers to the thickness change of the strut $i$ due to change in the tensile direction change $\Delta L_{1p}/L_0$.

Manipulation of (17) leads to:
\begin{equation}
\displaystyle \nu_{3}=\frac{1}{3}\left(-\frac{\Delta L_z^{(1)}/L_z^{(1)}}{\Delta L_{1p}/L_0}-2\frac{\Delta L_z^{(2)}/L_z^{(2)}}{\Delta L_{2p}/L_0}\frac{\Delta L_{2p}/L_0}{\Delta L_{1p}/L_0}\right),
\end{equation}
or 
\begin{equation}
\displaystyle \nu_{3}=\frac{1}{3}\left(\nu_{zig}+2\nu_{zig}\frac{\Delta L_{2p}/L_0}{\Delta L_{1p}/L_0}\right)=\frac{1}{3}\nu_{zig}\left(1+2\frac{\delta}{\epsilon}\right),
\end{equation}
which finally yields to

\begin{equation}
\displaystyle \nu_{3}=\nu_{zig}\;\frac{3+R}{3(1+R)}.
\end{equation}

Using the following definitions for the ratios between force constants $s \equiv k_t/k_s$ and $r \equiv k_\theta/k_t$, the Poisson's ratios can be expressed as

\begin{equation}
\displaystyle \nu_{1}=\frac{1-\beta}{3+\beta}
\label{eq-v10}
\end{equation} 
and
\begin{equation}
\displaystyle \nu_{3}=\frac{(1-s)(1+\beta)}{ (\tan^2\gamma + s)(3+\beta)},
\label{eq-v30}
\end{equation}
where $\beta=3k_t/k_{sb}= 3[1+(s-1)\cos^2 \gamma]/r$.

For the case where the nanotube struts have effectively infinite modulus ($s=0$), the expressions reduce to 
\begin{equation}
\displaystyle \nu_{1}=\frac{1-\beta}{3+\beta}
\label{eq-v1}
\end{equation} 
and
\begin{equation}
\displaystyle \nu_{3}=\frac{1+\beta}{3+\beta}\cot^{2}\gamma,
\label{eq-v3}
\end{equation}
but now with $\beta = (3 \sin^2\gamma)/r$. Using a different model, which includes a host of structural and force constant parameters in $\beta$, the above dependence of $\nu_1$ on $\beta$ has been predicted for sheets of cellulose-based papers \cite{perkins}.

\section{IV. Model results}
\subsection{Poisson's ratio behavior}
We can see from expressions (\ref{eq-v10}) and (\ref{eq-v30}) that when changes in the angle between intersecting nanotubes are negligible in comparison to changes in nanotube length (due to both nanotube stretching and changes in the angle $\theta$), expressed as $\beta \rightarrow \infty$, the most negative in-plane Poisson's ratio ($\nu_1\rightarrow -1$) and most positive thickness direction Poisson's ratio ($\nu_{3}\rightarrow \cot^{2}\gamma$) are obtained. On the other hand, when changes in the angle between intersecting nanotubes are much easier than changes in the nanotube length ($\beta \rightarrow 0$), $\nu_{1}\rightarrow 1/3$ and $\nu_{3}\rightarrow (1/3)\cot^{2}\gamma$. We can see that in both limits the thickness-direction Poisson's ratio is positive while the in-plane Poisson's ratio undergoes a transition from negative (auxetic) to positive (non-auxetic) values. 

This non-auxetic/auxetic behavior can be simply visualized  by noting that two neighboring nanotube layers in the Fig. 6(a) are coupled like the struts of a wine rack. If rotation between struts dominates, like for an ordinary wine rack, the Poisson's ratio is positive. If this torsional rotation of struts is blocked (by welding together the struts) and the struts are stretchable but not bendable, strut length increases produce a negative Poisson's ratio.

These results indicate that the present model allows a qualitative description of the experimental behavior observed for the carbon nanotube sheets \cite{hall} if the mixing of MWNTs in SWNT sheets can be represented by an effective change on the ratios $r$ and $s$ as MWNTs are incorporated. In order to estimate the utility of the model we can determine the $r$ and $\gamma$ values (considering the approximation $s=0$) that provide the observed values of Poisson's ratio (by matching the expressions (\ref{eq-v1}) and (\ref{eq-v3}) with the experimentally observed Poisson's ratios). The expressions (\ref{eq-v1}) and (\ref{eq-v3}) can be inverted yielding to

\begin{equation}
\displaystyle r=\frac{3(1 - \nu_{1}^2)}{( 3 \nu_{1} -1) (\nu_{1} -1 - 2 \nu_3)},
\label{eq_r}
\end{equation}
and
\begin{equation}
\displaystyle \cos\gamma= \sqrt{\frac{2 \nu_3}{1 - \nu_{1} + 2 \nu_3}}.
\label{eq_gamma}
\end{equation}

Using the experimental data for MWNT sheets ($\nu_{1}=-0.20$ and $\nu_{3}=0.75$) the derived values from the above expressions are $r\simeq0.67$ and $\gamma\simeq42^\circ$. Similarly, for SWNT sheets ($\nu_{1}=0.06$ and $\nu_{3}=0.33$) we have $r\simeq2.28$ and $\gamma\simeq50^\circ$. The predicted $\gamma$ values are consistent with average angles from x-ray diffraction of 41.7$^\circ$ for MWNT sheets and 45.0$^\circ$ for SWNT sheets \cite{hall}. With these values of $r$ and $\gamma$ we have that the ratio of the $\beta$ parameters for MWNT and SWNT sheets is $\beta_{\text{MWNT}}/\beta_{\text{SWNT}}\simeq2.6$.

\subsection{Negative linear compressibility}
Negative Poisson's ratios are sometimes accompanied by much rarer mechanical properties: negative linear compressibilities and negative area compressibility, meaning that a material expands in either one or two orthogonal directions when hydrostatic pressure is applied \cite{baugh}. A negative linear compressibility is the inverse of another strange property - increasing density when elongated in a direction where linear compressibility is negative, and both require that $1-\nu_1-\nu_3 < 0$ \cite{baugh}. Using (\ref{eq_r}) and (\ref{eq_gamma}) we have that this condition becomes

\begin{equation}
\displaystyle 1-\nu_{1}-\nu_{3}=\frac{2 r-3 \cos ^2\gamma -r \cot ^2\gamma+6 \sin ^2\gamma }{3(r+ \sin ^2\gamma)}.
\end{equation}

Thus negative in-plane compressibility, negative area compressibility for the sheet plane, and stretch densification is predicted for $\cos \gamma > \sqrt{2/3}$, which implies $\gamma < 35.3^o$. Fig. 8 shows the predicted behavior of $\nu_{1}$ against $\nu_{3}$ for the pure SWNT ($r=2.28$) and MWNT ($r=0.67$) samples for the case where $s=0$. The average $\gamma$ needed for achieving these properties will decrease as a result of in-plane nanofiber meandering, since only the tensile strain component resulting in thickness change effects $\nu_3$. Since the enhanced degree of in-plane alignment needed to realize negative linear compressibilities is not large,  the needed improvement on the degree of in-plane alignment might be obtainable by either using high pressure for the filtration step used for sheet fabrication or by annealing the as-fabricated sheets under high mechanical load.
\vspace{0.5cm}
\begin{center}
\begin{figure}[h]
\includegraphics[angle=0,scale=0.3]{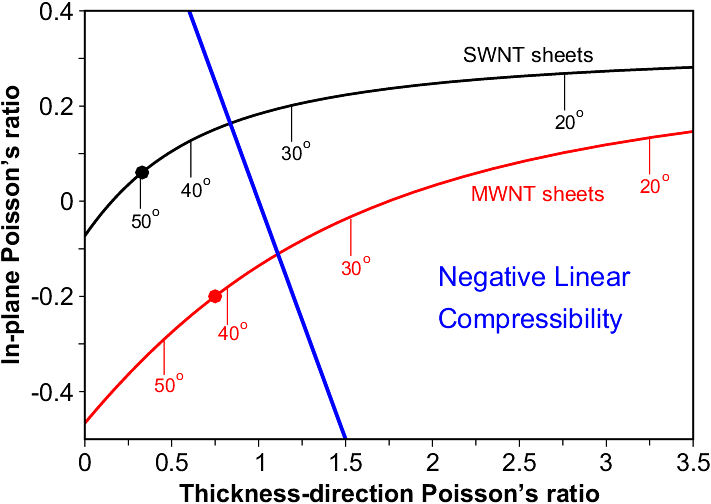}
\caption{(Color online) The relationships between in-plane and thickness-direction Poisson's ratios as a function of indicated average nanotube orientation angle $\gamma$ for SWNT and MWNT sheets having the force constant ratios $r$ that yield the measured Poisson's ratios (circles). The straight (blue) line represents the equation $1-\nu_1-\nu_3=0$. Values of $\gamma$ on the right side of this line are predicted to produce negative linear compressibilities for SWNT and MWNT sheets.}
\end{figure}
\end{center}

\vspace{-1.0cm}
\section{V. Discussions}
In order to discuss the role of the nanoscale constituents of nanotube sheets on the behavior of the in-plane Poisson's ratios we will proceed as follows. Firstly, we will show that beam bending in response to tensile stress presents the same effect for increasing Poisson's ratio as torsional rotation. This allows the use of the force constant ratio between strut bending to stretching for understanding the in-plane Poisson's ratio values for nanotube sheets. Secondly, we will estimate this ratio for the nanotube sheets based on experimentally accessible quantities and use this result for predicting the in-plane Poisson's ratios for SWNT and MWNT buckypaper.

\subsection{Nanotube bending vs. nanotube stretching}
In analogy with a wine rack, positive in-plane Poisson's ratios would result if (i) the hinges are welded to struts to prohibit torsional rotation and (ii) the struts are much easier to bend than to stretch. Nanotube beam bending in response to a tensile stress within the sheet plane changes the effective angle between intersecting nanotubes, and produces a corresponding increase in Poisson's ratio, similarly to the response when there are changes in torsional angle for the model of Fig. 6(a). If fiber beam bending is the predominant deformation that changes the effective angle between intersecting fibers and fiber deviation from in-plane orientation is neglected ($\gamma=0$), $\beta = 3k_b/k_{sb}$, where the force constants for fiber bending and tensile fiber elongation are $k_b$ and $k_{sb}$, respectively.

In order to obtain the dependence of $\nu_1$ on $k_b/k_{sb}$ we proceed as follows. Since only the in-plane Poisson's ratio $\nu_1$ is presently being evaluated, we used a two-dimensional (2D) sheet structure to represent the dependence of $\nu_1$ on effective $k_b/k_{sb}$. These 2D sheets look like the projection of the structure shown in Fig. 6(a), except that six co-planar struts meet in 2D at each junction.  With the purpose of excluding torsional angle changes, but enabling strut length changes and strut bending, the struts were represented by long chains composed of $N$ ``atoms''($N$ large), which were allowed to undergo bond angle bending at each atom and elongation of the bonds between atoms (see Fig. 9). The struts were connected to artificially contracted six membered rings that were so small and so rigid with respect to angle and dimensional changes that they acted as an junction that does not allow the equivalent of torsional rotation. For a single value of $k_b/k_{sb}$ the structure was geometrically optimized and the Poisson's ratio was obtained from the stress tensor derived from the second derivative of the energy with respect to the strain. These calculations were carried out using the Cerius$^2$ open force field molecular mechanics \cite{cerius,details}. By varying strut bending and strut elongation moduli arbitrarily, the calculations provided $\nu_1 = (1-\beta)/(3+\beta)$, with $\beta = 3 k_b/k_{sb}$ as shown in Fig. 10 \cite{details}. An illustration of the deformations presented by the 2D model under tensile strain is depicted in Fig. 11.

\begin{center}
\begin{figure}[h]
\includegraphics[angle=0,scale=0.3]{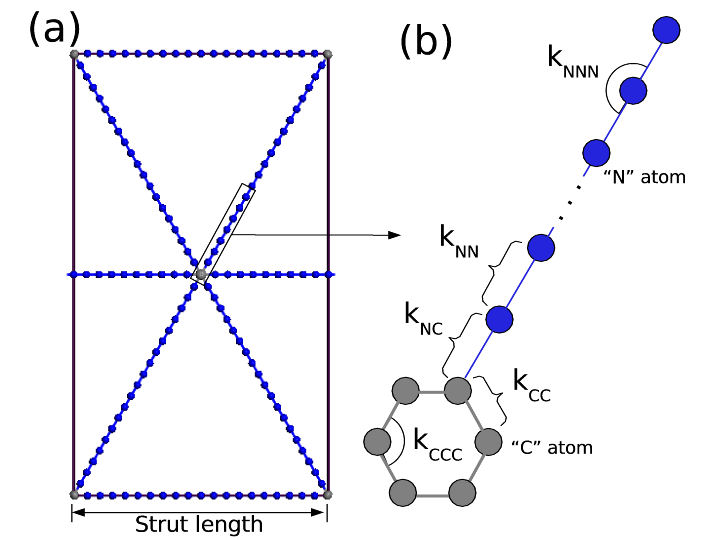}
\caption{(Color online) Schematic representation of the 2D sheets used in molecular mechanics calculations. (a) Representation of the unit cell. (b) Detailed view of a region of (a) showing the main spring constants used to describe the 2D deformations \cite{details}.}
\end{figure}
\end{center}

\begin{center}
\begin{figure}[h]
\vspace{0.5 cm}
\includegraphics[angle=0,scale=0.3]{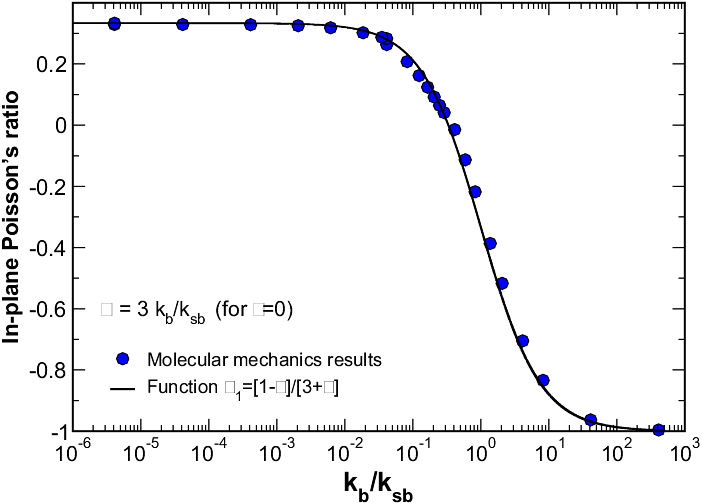}
\caption{(Color online) In-plane Poisson's ratio vs. $k_b/k_{sb}$ obtained from molecular mechanics calculations.}
\end{figure}
\end{center}

\begin{center}
\begin{figure}[b]
\includegraphics[angle=0,scale=0.3]{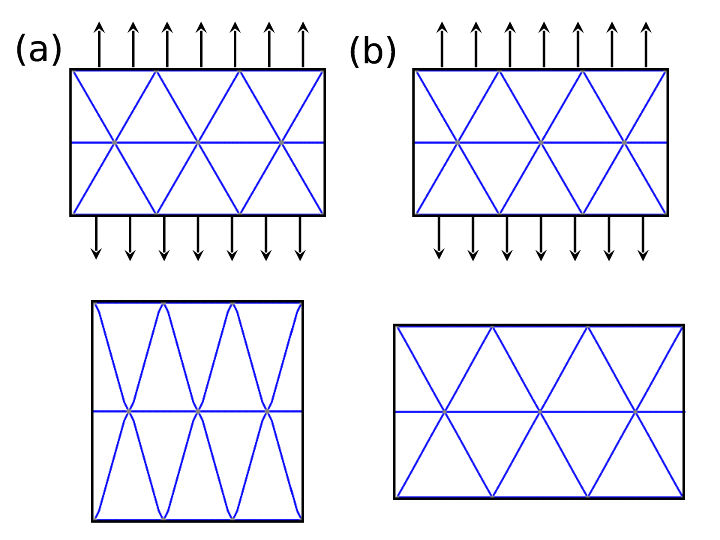}
\caption{(Color online) Resulting deformations of the 2D sheet model under tensile strain for (a) $k_s/k_{sb}\rightarrow 0$ (non-auxetic) and (b) $k_s/k_{sb}\rightarrow \infty$ (auxetic). Top (bottom) diagrams correspond to the underformed (deformed) state.}
\end{figure}
\end{center}

\vspace{-2.0cm}
\subsection{Nanoscale origin of the auxetic property of nanotube sheets}
Having shown that beam bending has a similar effect for increasing in-plane Poisson's ratio as torsional rotation, indicating that the ratio $k_b/k_{sb}$ plays an important role in the understanding of the in-plane Poisson's ratio behavior, we will estimate $k_b/k_{sb}$ for the nanotube sheets. Before that, it is worth mentioning that when $\gamma=0$ ($\theta=90^\circ$) and changes in the $\theta$ angle are negligible compared to changes in the strut lengths ($k_\theta/k_s\rightarrow \infty$) $k_{sb}\rightarrow k_s$. Thus, the differences between MWNTs and SWNT bundles will be associated with the intrinsic mechanical properties of each of them, being represented here by effective spring constants related to nanotube axial stretching $k_s$ and to the nanotube lateral bending $k_b$. Therefore, the ratio $k_b/k_s$ will be different for each nanotube type and will depend upon their nanoscale structure.  

Consider a perfect, straight SWNT of radius $r_t$ having length $l$ described by the elastic tube model \cite{dibiasio,elastic}. In this case, $k_s = Y_s A/l$ and $k_b = 3 Y_b I/l^3$, where $Y_s$ and $Y_b$ are the Young's modulus for stretching and bending, respectively, $A$ is its cross-section area, and $I$ is the moment of inertia.  Using $A=2\pi r_t h$, $Y_s= C/h$ and  $Y_b I = K = \pi C r_t^3$ \cite{boris}, where $h$ is the graphene sheet thickness, we obtain $k_b/k_s=(3/2)(r_t/l)^2$. Here we assumed that bending and stretching Young's modulus have the same value.

For a bundle  of radius $R_b$ composed by $p$ SWNTs of radius $r_t$ that can freely slip along each other, acting independently, the bending force constant is expressed as $p k_b$. Assuming that the load is carried by the SWNTs on the perimeter of each bundle \cite{ruoff}, the stretching force constant will be given by $k_s=(C/h) (A_{\text{ext}}/l)$, where $A_{\text{ext}}$ is the cross section area of the outer nanotubes in the SWNT bundle. For large diameter bundles ($R_b\gg r_t$) $A_{\text{ext}}\simeq 2\pi R_b (2r_t)$. For SWNTs closely packed into a two-dimensional hexagonal array, the volume fraction $V_p$ is 0.906 thus $p=V_b(\pi R_b^2)/(\pi r_t^2) = V_b(R_b/r_t)^2$. Therefore,

\begin{equation}
\displaystyle
\left(\frac{k_b}{k_s}\right)_{\text{bundle}} = \frac{3hV_b}{4}\left(\frac{R_b}{l^2}\right).
\label{eq-swnt}
\end{equation}

The bending stiffness of a MWNT can be expressed simply as a sum of bending stiffness of the individual and independent SWNTs \cite{boris}, i.e., $K=\pi \sum_{i=n_i}^{n_f}{C_i r_{t(i)}^3}$, where $i$ is the corresponding wall number of the MWNT. For a large diameter MWNT it is reasonable to assume that all $C_i$ for this MWNT are the same and equal to the in-plane stiffness $C$. Considering that the MWNT is composed only by armchair SWNTs then  $r_{t(i)}= \alpha i$, $\alpha=0.68$ {\AA} \cite{dresse} thus $K=\pi \xi  C \alpha^3$, where $\xi=\sum_{i=n_i}^{n_f}{i^3}$. Assuming that the load is supported by only the outer wall of the MWNT then $k_s=(C/h)(A_{\text{outer}}/l)$ where $A_{\text{outer}}=2\pi r_{t(n_f)}h$. Consequently,
\begin{equation}
\displaystyle
\left(\frac{k_b}{k_s}\right)_{\text{MWNT}} = \frac{3\alpha^3}{2}\left(\frac{\xi }{l^2 \;r_{t(n_f)}}\right).
\label{eq-mwnt}
\end{equation}

From eqs. (\ref{eq-swnt}) and (\ref{eq-mwnt}) we can see that to decrease the in-plane Poisson's ratio (through the increase $\beta=3k_b/k_s$), inter-junction lengths $l$ should decrease or, equivalently, sheet density should increase. For SWNT sheets, increasing bundle radius also leads to a decrease of the Poisson's ratio. For the limiting case where the bundle is formed by only one SWNT, $k_b/k_s=(3/2)(r_t/l)^2$. The largest geometrically possible value of $r_t/l$ is $\sin(60^\circ)/2$, which corresponds to the physically unreasonable case where each layer within the nanotube sheet comprises straight nanotubes that are close packed within the layer. From (23) and $\beta=3k_b/k_s$, this hypothetical buckypaper of perfectly straight, infinitely long, unbundled SWNTs can not have a Poisson's ratio below 0.04. Since $\beta$ for buckypaper-like sheets comprising long circular solid fiber is the same as for a SWNT when the effective Young's modulus for bending equals that for tension, the predicted $\nu_1$ is also 0.04 or higher, and likely much higher since buckypaper-like sheets or ordinary paper do not have fibers that are close packed in a plane.

For MWNT sheets, a decrease of the in-plane Poisson's ratio can be obtained by increasing the number of interior walls (which increases $\xi$) and the MWNT diameter. While all nanotube walls contribute additively to $k_b$, only the outer wall contributes to $k_s$ unless the MWNTs are extremely long. However, the effects of these structure changes are not simple, since increasing $k_b$ and decreasing $l$ can decrease nanotube meandering between junctions and this decrease of meandering can provide a positive contribution to $k_s$. 

In order to have a rough estimate of $k_b/k_s$ for MWNTs and SWNT bundles present in the nanotube sheets (Fig. 2), we use experimental data \cite{hall} showing that the SWNTs have an average diameter of about 1.0 nm and a average bundle diameter of 20 nm and that the MWNTs have an outer diameter of about 12 nm, and contain about nine walls (Fig. 12). From these data $n_i=48$, $n_f=88$, $\xi=3135888$, and $k_b/k_s \simeq 2/l^2$ and $k_b/k_s \simeq 247/l^2$ ($l$ in nm) for SWNT and MWNT sheets, respectively. For MWNTs and SWNT bundles sharing approximately the same length we can see that $\beta_{\text{MWNT}}/\beta_{\text{SWNT}}\sim 100$, two orders of magnitude larger than the previously obtained value of 2.6 when the nanotubes are considered completely rigid. While these estimates lead to positive and negative values for the in-plane Poisson's ratio for SWNT and MWNT sheets, respectively, in qualitatively agreement with experiment, quantitative agreement is poor (0.3 and $-$0.6 for SWNT and MWNT sheets, respectively, with $l\sim 10$ nm). As we will see, this difference lies on the assumption that the effective force constant for elongating the nanotubes corresponds to the modulus of an individual straight nanotube. Because of nanotube meandering, and possibly elasticity at inter-nanotube junctions, this assumption is not valid.

In order to better estimate $\nu_1$ we can proceed as follows. The Young's modulus of the nanotube sheet is obtained by taking the second derivative of the per-strut energy of (11) with respect to $\epsilon$, 

\begin{equation}
\displaystyle
Y= \frac{L_0 k_{sb}(1-\nu_1)}{2V_s},
\label{eq_Y}
\end{equation}
where we used $R = k_{sb}/k_t = 3(1+\nu_1 )/(1-3\nu_1)$ from (15) and $V_s$ represents the volume per strut in the sheet structure. With this expression and the experimentally observed Young's modulus for the nanotube sheets, we can obtain a more precise estimate of $k_{sb}$, instead of only $k_s$, as previously considered in the limiting case of $\gamma=0$.

Deriving the effective strut lengths for the highly disordered SWNT and MWNT buckypaper is important, since $L_0$ and $V_s$ are needed for comparing theory with experiment. In order to calculate these parameters, we look at the intersecting nanotubes (or nanotube bundles) as being stacked in the thickness direction like layers of logs having an effective diameter $D$, where $D$ is the sum of the covalent diameter of the nanofiber and the 0.34 nm van der Waals diameter of carbon. Correspondingly, the volume per strut is $D L_0^2 \sin(120^\circ)$. If the strut weight per strut length is $W_L$ and the measured nanotube sheet density is $\rho$, then $\rho = W_L/[D L_0^2 \sin(120^\circ)]$. Using the observed densities for the MWNT (0.343 g/cm$^3$) and SWNT sheets (0.692 g/cm$^3$) \cite{hall}, the corresponding calculated $L_0$ for MWNT and SWNT struts are 54.3 nm and 39.5 nm, respectively. While these distances seem shorter than suggested by the micrographs of Fig. 2, note that these micrographs are for the sheet surface (the face originally in contact with the filter membrane) and do not provide the junction density and corresponding $L_0$ in the buckypaper interior. Using this strut volume, eq. (\ref{eq_Y}) becomes 
\begin{equation}
\displaystyle
k_{sb}= \frac{2YD\sin(120^\circ)}{1-\nu_1}.
\end{equation}

The $k_b$ for the MWNTs is obtained by the sum of bending force constants for all component SWNTs (with $C=345$ J/m$^2$ \cite{boris}) and the $k_b$ for SWNT bundles is derived from the measured average Young's modulus for bending ($Y_b$) 20 nm diameter SWNT bundles (50 GPa) \cite{kis}, using the force constant for bending a solid cylindrical rod $k_b = 3\pi r_t^4 Y_b/(4l^3)$ \cite{dibiasio}. Using the observed Young's modulus (1.81 GPa and 3.21 GPa for MWNT and SWNT sheets, respectively) $\beta$ (=$k_b/k_{sb}$) can be self-consistently obtained from the relations above, which provides 1.84 and 0.42 for MWNT and SWNT sheets, respectively. Therefore, the predicted in-plane Poisson's ratios are $-$0.17 for MWNT buckypaper (vs. the observed $-$0.20) and 0.17 for SWNT buckypaper (vs. the observed 0.06). Considering $Y_b = 81$ GPa, which is  within the range of experimental uncertainty \cite{kis}, the calculated $\nu_1$ for SWNT buckypaper assumes the observed value. For this latter estimate we can see that $\beta_{\text{MWNT}}/\beta_{\text{SWNT}}\simeq2.4$ in agreement with the previously predicted value of 2.6 for the case where spring constant for nanotube stretching is much larger than strut torsional rotations ($s = k_t/k_s \rightarrow 0$), fitted from observed in-plane and thickness direction Poisson's ratios.

\begin{center}
\begin{figure}[h]
\includegraphics[angle=0,scale=0.3]{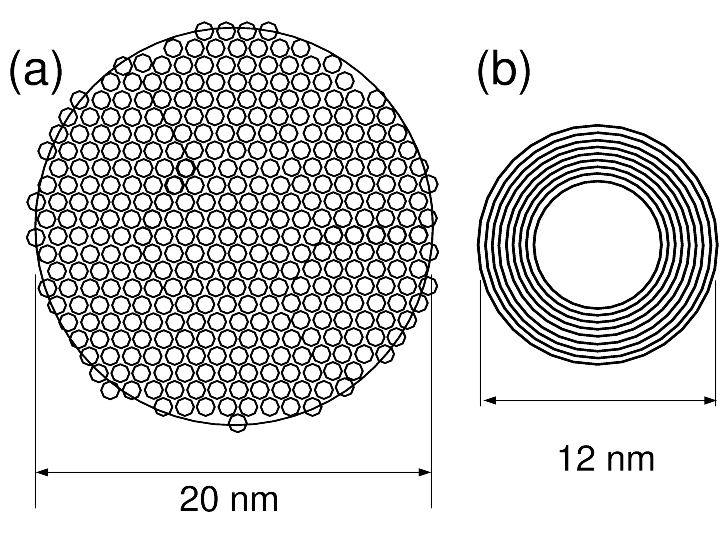}
\caption{Schematic representation of the (a) SWNTs and (b) MWNTs present in nanotube sheets (Fig. 2).}
\end{figure}
\end{center}

\vspace{-1.0cm}
Having predicted the Poisson's ratios for the MWNT content limiting cases of 0 wt. \% (SWNT sheets) and 100 wt. \% (MWNT sheets) we expect that MWNT/SWNT mixing in buckypapers can be represented by altering the intrinsic characteristics of the fibers (nanotube bending and elongation force constants) as well as by modifying the structural morphology (sheet density, fiber connections) to provide the observed intermediate Poisson's ratios values. Further investigations are necessary to explore all these possibilities and thus to provide a better understanding of the behavior of the Poisson's ratios for buckypapers having intermediated MWNT content. 

\section{VI. Summary and Conclusions}
We proposed a relatively simple model to represent the complex morphology of carbon nanotube sheets and to describe the behavior of the Poisson's ratios. The model incorporates key structural features of the nanotube sheets: isotropic in-plane mechanical properties, nanotubes preferentially oriented in the sheet plane, and freedom to undergo stress-induced elongation as a result of straightening meandering nanotubes and changing the angle between intersecting nanotubes. The nanoscale origin of the constituent elements of the nanotube sheets is shown to play a fundamental role in determining the mechanical behavior that leads to auxetic features, specially the ratio between the force constants associated with the fiber bending and elongation. Qualitative and quantitative agreement with experiment were obtained using the Poisson's ratio expressions derived from the model, encouraging its use in predicting future results and new properties. Specifically, negative linear compressibility (material expansion in the sheet plane under hydrostatically compression) was predicted for carbon nanotube sheets with the average angle of the fibers with respect to the sheet plane  smaller than about 35$^\circ$, a value which is not too far from already produced nanotube sheets.

\begin{acknowledgments}
This work was supported by National Science Foundation grant DMI-0609115, Air Force Office of Scientific Research grant FA9550-05-C-0088, Lintec Corporation, and the Brazilian agencies FAPESP, FAPEMIG, Capes, and CNPq.
\end{acknowledgments}


\begin{thebibliography}{99}
\bibitem{lakes}R. S. Lakes, \textit{Science} {\bf 235}, 1038 (1987).

\bibitem{gibson} L. J. Gibson, M. F. Ashby, \textit{Proc. R. Soc. London A} {\bf 382}, 43 (1982).
 
\bibitem{evans}K. E. Evans, M. A. Nkansah, I. J. Hutchinson, S. C. Rogers, \textit{Nature} {\bf 353}, 124 (1991).

\bibitem{milton}G. Milton, \textit{J. Mech. Phys. Solids} {\bf 40}, 1105 (1992).

\bibitem{2}K. E. Evans, A. Alderson, \textit{Advanced Materials} {\bf 12}, 617 (2000).

\bibitem{baughman}R.H. Baughman, \textit{Nature} {\bf 425}, 667 (2003).

\bibitem{3}W. Yang, Z. Li, W. Shi, B. Xie, M. Yang, \textit{J. Mat. Sci.} {\bf 39}, 3269 (2004).

\bibitem{doug}R.H. Baughman, D. S. Galv\~ao, \textit{Nature} {\bf 365}, 735 (1993).

\bibitem{hall} L. J. Hall, V. R. Coluci, D. S. Galv\~ao, M. E. Kozlov, M. Zhang, S. O. Dantas, R. H. Baughman, \textit{Science} \textbf{320} 504 (2008).

\bibitem{popov}V. N. Popov, V. E. Van Doren, M. Balkanski, \textit{Phys. Rev. B} \textbf{61}, 3078 (2000).

\bibitem{shen} L. Shen, J. Li, \textit{Phys. Rev. B} \textbf{69}, 045414 (2004).

\bibitem{mintmire}H. M. Lawler, J. W. Mintmire, C. T. White, \textit{Phys. Rev. B} \textbf{74}, 125415 (2006).

\bibitem{elastic}B. I. Yakobson, C. J. Brabec,  J. Bernholc, \textit{Phys. Rev. Lett.} \textbf{76}, 2511 (1996).

\bibitem{lu} J. P. Lu, \textit{Phys. Rev. Lett.} \textbf{79}, 1297 (1997).

\bibitem{hernandez} E. Hernandez, C. Goze, P. Bernier, A. Rubio, \textit{Phys. Rev. Lett.} \textbf{80}, 4502 (1998).

\bibitem{portal}D. Sanchez-Portal, E. Artacho, J. M. Soler, A. Rubio, P. Ordejon,  \textit{Phys. Rev. B} \textbf{59}, 12678 (1999).

\bibitem{vgraf1} O. L. Blakslee, D. G. Proctor, E. J. Seldin, G. B. Spence, T. Weng, \textit{J. Appl. Phys.}  \textbf{41}, 3373 (1970).

\bibitem{vgraf2} J. Seldin, C. W. Nezbeda, \textit{J. Appl. Phys.}  \textbf{41}, 3389 (1970).

\bibitem{berhan} L. Berhan, Y. B. Yi, A. M. Sastry, E. Munoz, M. Selvidge, R. Baughman, \textit{J. Appl. Phys.} \textbf{95}, 4335 (2004).

\bibitem{cox} H. L. Cox, \textit{Br. J. Appl. Phys.} \textbf{3}, 72 (1952).

\bibitem{warren} W. E. Warren, A. M. Kraynik, \textit{Mechanics of Materials} \textbf{6}, 27 (1987).

\bibitem{narter} M. A. Narter, S. K. Batra, R. R. Buchanan, \textit{Proc. R. Soc. Lond. A} \textbf{455}, 3543 (1999).

\bibitem{astrom} J. A. {\AA}strom, J. P. Makinen, M. J. Alava, J. Timonen, \textit{Phys. Rev. E} \textbf{61}, 5550 (2000).

\bibitem{wu} X.-F. Wu, Y. A. Dzenis, \textit{J. Appl. Phys.} \textbf{98}, 093501 (2005).

\bibitem{peter} P. V. Pikhitsa, \textit{Phys. Rev. Lett.} \textbf{93}, 015505 (2004).

\bibitem{zhang}M. Zhang, K. R. Atkinson, R. H. Baughman, \textit{Science} {\bf 306}, 1358 (2004).

\bibitem{yudasaka}M. Yudasaka, H. Kataura, T. Ichihashi, L. C. Qin, S. Kar, S. Iijima, \textit{Nano Lett.} {\bf 1}, 487 (2001).

\bibitem{bachilo}S. M. Bachilo, M. S. Strano, C. Kittrell, R.H. Hauge, R. E. Smalley, R. B. Weisman, \textit{Science} {\bf 298}, 2361 (2002).

\bibitem{nikolaev}P. Nikolaev, M. J. Bronikowski, R. K. Bradley, F. Rohmund, D. T. Colbert, K. A. Smith, R. E. Smalley, \textit{Chem. Phys. Lett.} {\bf 313}, 91 (1999).

\bibitem{rinzler}A. G. Rinzler, J. Liu, H. Dai, P. Nikolaev, C. B. Huffman, F. J. Rodriguez-Macias, P. J. Boul, A. H. Lu, D. Heymann, D. T. Colbert, R. S. Lee, J. E. Fischer, A. M. Rao, P. C. Eklund, R. E. Smalley, \textit{Appl. Phys. A} {\bf 67}, 29 (1998).

\bibitem{ran}S. Ran, D. Fang, X. Zong, B. S. Hsiao, B. Chu, P. M. Cunniff, \textit{Polymer} {\bf 42}, 1601 (2001).

\bibitem{landau} L. D. Landau and E. M. Lifshitz, \textit{Theory of Elasticity}, 3rd Edition.

\bibitem{boris}B. I. Yakobson, L. S. Couchman, \textit{J. Nanoparticle Research} \textbf{8}, 105 (2006).

\bibitem{dibiasio}C. M. DiBiasio, M. A. Cullinan, M. L. Culpepper, \textit{Appl. Phys. Lett.} \textbf{90}, 203116 (2007).

\bibitem{egg} J. N. Grima, J. J. Williams, K. E. Evans, \textit{Chem. Comm.}  4065 (2005).

\bibitem{delince}M. Delinc\'e, F. Delannay, \textit{Acta Mater.} \textbf{52}, 1013 (2004).

\bibitem{baugh}R. H. Baughman, S. Stafstr\"om, C. Cui, S. O. Dantas, \textit{Science} \textbf{279}, 1522 (1998).

\bibitem{perkins}R. W. Perkins, in \textit{Proceedings of the Conference on Paper Science and Technology $-$ The Cutting Edge: Fiftieth Anniversary Year 1929-1979} (Institute of Paper Chemistry, Appleton, WI, 1980), VII, pp. 89-111.

\bibitem{cerius}
{http://www.accelrys.com/cerius2/}.

\bibitem{details}The 2D sheet model presents Pmmm symmetry with $a$=101 {\AA}, $b$=175 {\AA}, $c$=300 {\AA}, and 126 ``atoms'' in the unit cell. Each ``atom'' represents a junction between bonds. ``C''-type atoms were used in the six-membered ring and ``N''-type atoms composed the strut. The equilibrium ``bond'' distances between ``C''-atoms was 0.5 {\AA} and between ``N''- and ``C''-atoms and ``N''-atoms was 5.0 {\AA}. The strut is composed by 20 bonds having a total length of 101 {\AA}. The bond stretching spring constants were $k_{\text{CC}}=k_{\text{CN}}$=1000 kcal/mol.{\AA}$^2$ and the angle bending spring constants  $k_{\text{CCC}}=k_{\text{CCN}}=k_{\text{NNC}}$ = 10000 kcal/mol.rad$^2$. The spring constants $k_{\text{NNN}}$ and $k_{\text{NN}}$ were arbitrarily varied to obtain the results presented in Fig. 10. 

\bibitem{ruoff}M. F. Yu, B. S. Files, S. Arepalli, R. S. Ruoff, \textit{Phys. Rev. Lett.} \textbf{84}, 5552 (2000).

\bibitem{dresse} R. Saito,  G. Dresselhaus, M. S. Dresselhaus  
\textit{ Physical Properties of Carbon Nanotubes}, Imperial College Press, London (1998).

\bibitem{kis} A. Kis, G. Csanyi, J. P. Salvetat, T. N. Lee, E. Couteau, A. J. Kulik, W. Benoit, J. Brugger, L. Forr\'o, \textit{Nature Mat.} \textbf{3}, 153 (2004).
 
\end{thebibliography}
\end{document}